\documentclass{aa}

\usepackage{graphicx}
\usepackage{amssymb}
\usepackage{amsmath}
\usepackage{natbib}
\bibpunct{(}{)}{;}{a}{}{,}
\def\puncspace{\ifmmode\,\else{\ifcat.\C{\if.\C\else\if,\C\else\if?\C\else%
\if:\C\else\if;\C\else\if-\C\else\if)\C\else\if/\C\else\if]\C\else\if'\C%
\else\space\fi\fi\fi\fi\fi\fi\fi\fi\fi\fi}%
\else\if\empty\C\else\if\space\C\else\space\fi\fi\fi}\fi}
\def\SP{\let\\=\empty\futurelet\C\puncspace}
\def\etal{{\it et al.\/}\ }
\def\eg{{\it e.g.\/}\rm,\ }
\def\void#1{{}}
\def\lsim{\lesssim}
\def\gsim{\gtrsim}
\def\kms{km~s$^{-1}$\SP}

\begin{document}

\title{Optically-selected clusters at $0.8\lsim z \lsim 1.3$ in the EIS Cluster
Survey}

\author{C. Benoist\inst{1} \and L. da Costa\inst{2} \and
H.E. J{\o}rgensen\inst{3} \and L.F. Olsen\inst{3} \and S. Bardelli\inst{4} \and 
E. Zucca\inst{4} \and M. Scodeggio\inst{5}
\and D. Neumann\inst{6} \and M. Arnaud\inst{6} \and S. Arnouts\inst{2}
\and A. Biviano\inst{7} \and M. Ramella\inst{7}}

\institute{
     Observatoire de la C\^ote d'Azur, CERGA, 
     BP 229, 06304 Nice, cedex 4, France 
\and European Southern Observatory, Karl-Schwarzschild-Str. 2,
     D-85748 Garching b. M\"unchen, Germany
\and Astronomical Observatory, University of Copenhagen,
     Juliane Maries Vej 30, DK-2100 Copenhagen, Denmark
\and INAF - Osservatorio Astronomico di Bologna, 
     via Ranzani 1, I-40127 Bologna, Italy
\and Istituto di Fisica Cosmica - CNR, Milano, Italy
\and SAp, CEA/Saclay, L'Orme des Merisiers Bat. 709, 
     91191 Gif-sur-Yvette Cedex, France 
\and INAF - Osservatorio Astronomico di Trieste, Via G. B. Tiepolo 11,
     I-34131 Trieste, Italy}

\date{Received \today ; accepted}

\abstract{This paper presents preliminary results of a spectroscopic
survey being conducted at the VLT of fields with optically-selected
cluster candidates identified in the EIS $I$-band survey. Here we
report our findings for three candidates selected for having estimated
redshifts in the range $z=0.8-1.1$. New multi-band optical/infrared
data were used to assign photometric redshifts to galaxies in the
cluster fields and to select possible cluster members in preparation
of the spectroscopic observations. Based on the available
spectroscopic data, which includes 147 new redshifts for galaxies with
$I_{AB}\lsim22-23$, we confirm the detection of four density
enhancements at a confidence level $>99\%$.  The detected
concentrations include systems with redshifts $z=0.81$, $z=0.95$,
$z=1.14$ and the discovery of the first optically-selected cluster at
$z=1.3$.  The latter system, with three concordant redshifts,
coincides remarkably well with the location of a firm X-ray detection
($>5\sigma$) in a $\sim80$~ksec XMM-Newton image taken as part of this
program which will be presented in a future paper (Neumann \etal
2002). The $z>1$ systems presented here are possibly the most distant
identified so far by their optical properties alone.
\keywords{Galaxies: clusters: individual: EIS0046-2930, EIS0533-2412,
EIS0954-2023 -- large-scale structure of Universe -- Cosmology:
observations}}

\maketitle

\section{Introduction}
\label{sec:intro}

Clusters of galaxies are both ideal sites for studying galaxy
evolution and important cosmological probes, especially at redshifts
$z\gsim0.5$, where differences between competing evolutionary and
cosmological models become important. This has motivated several
searches for distant clusters using a variety of techniques in
different wavelengths. As a result, over the past few years a
remarkable progress has been made in detecting an ever increasing
number of systems with $z \gsim0.5$ (see Gioia 2000 for a recent
review). More recently, a handful of clusters at $z\gsim1$ have also
been identified. While the sheer existence of these high redshift
clusters is of great importance, the current number of confirmed
systems is still very small, mostly identified from serendipitous
X-ray searches (\eg Rosati \etal 1999) or from infra-red imaging
(Stanford \etal 1997).  Therefore, the construction of a large
sample of confirmed clusters at $z\gsim0.8$ representative of the
entire population of these high-z systems remains an important goal of
observational cosmology. However, as these systems are expected to be
rare, finding them requires large areas of the sky to be covered,
limiting the techniques that can be used in identifying candidates. In
particular, surveys at X-ray and mm wavelengths (Carlstrom \etal 2000)
are unlikely to provide in the near future the necessary sky coverage
for constructing the large samples of very distant clusters of
galaxies required for statistical analyses.

An alternative way is to consider multi-band optical/infrared imaging
data. Thanks to the advent of panoramic CCD imagers, wide-angle
imaging surveys in the optical and near infrared wavelengths have
become viable and can be used for identifying cluster candidates up to
$z\sim1$. Examples of wide-angle surveys that have been used to
identify intermediate to high redshift clusters include those of Gunn
\etal (1986), Postman \etal (1996), the ESO Imaging Survey (EIS)
Cluster Survey (Olsen \etal 1999a,b; Scodeggio \etal 1999), the
Red-Sequence Survey (Gladders \& Yee 2000) and the Las Campanas
distant cluster survey (Gonzalez \etal 2001).  These surveys,
especially those carried out in a single passband, can only provide
plausible candidates and further benefits from additional
multi-wavelength observations to mitigate many of the problems of
foreground-background contamination, to assign photometric redshifts
for galaxies of different morphological types and to select possible
cluster members to improve the yield of spectroscopic follow-ups.

In this paper we describe our first attempts to explore the nature of
the high redshift cluster candidates identified in the EIS $I$-band
survey, combining new imaging and spectroscopic observations.
Altogether there are about 82 candidates with matched-filter redshifts
$\gsim 0.8$ for which about half have already been complemented by
imaging observations in $BVRJK$. Among the various clusters for which
we have spectroscopic data, we present here three clusters at higher
redshift.  In section~\ref{sec:data} we describe the selection of the
candidate clusters and of the galaxy sample used in the
observations. In section~\ref{sec:obs_red}, we briefly describe the
reduction procedure which will be expanded in a separate paper where
the accumulated data are presented (J{\o}rgensen \etal 2002). In
section~\ref{results}, the observed redshift distribution and the
technique used to identify groups in redshift space are
presented. Finally, in section~\ref{summary} our main results are
summarised.

\section{Cluster and Galaxy Sample}
\label{sec:data}

The results presented here are part of an ongoing comprehensive effort
to identify and study clusters at different epochs using as a starting
point the EIS cluster candidate compilation. This sample, consisting
of over 300 candidates, has been split roughly into three redshift
domains - low ($z\lsim0.4$), intermediate ($0.4\lsim z \lsim 0.7$) and
high ($z\gsim 0.7$). Several photometric and spectroscopic follow-up
programs are underway at different facilities to secure the necessary
data for confirmation (\eg Olsen \etal 2001) and more detailed
studies. The observations include moderately deep optical/infrared
imaging in $R$ and $JK$ (Scodeggio \etal 2002), spectroscopic
observations of intermediate redshift clusters at the ESO 3.6m
telescope (Ramella \etal 2000, Biviano \etal 2002), deep multi-band
imaging (Schirmer \etal 2002) for cosmic shear analysis and
spectroscopic observations of high redshift candidates at the VLT, and
for one case XMM-Newton data (Neumann \etal 2002). In the present
paper we focus our attention on three high-redshift ($z\gsim0.8$)
candidates - EIS0046-2930, EIS0954-2023 and EIS0533-2412 (Olsen \etal
1999b; Scodeggio \etal 1999)- for which photometric, spectroscopic and,
in one case, X-ray data are available.

The sample of objects selected for the spectroscopic observations in
the fields considered were drawn from an area of $10 \times 10$~arcmin
centred at the position of the candidate clusters as determined by the
matched-filter analysis of the $I$-band data. For EIS0046-2930 and
EIS0954-2023 the full area is covered in $BVI$, from publicly
available EIS-WIDE and/or EIS-Pilot data (Nonino \etal 1999;
Benoist \etal 1999), while the central $5\times5$~arcmin area is also
covered in $JK$ from SOFI observations at the NTT. While we now also
have $R$-band images these were not available at the time these
observations were being prepared. For these clusters the targets were
selected using one of the following criteria: {\it i)} using the
photometric redshifts computed in the area covered by the infrared
data (\eg Arnouts \etal 1999); {\it ii)} searching for the expected
$(I-K)$ and $(J-K)$ colours of early-type galaxies in the redshift
interval of interest; {\it iii)} identifying B- and V-dropouts
(expected for early-type galaxies considering the depth of EIS-WIDE)
in the outer part of the field; and {\it iv)} arbitrarily to fill the
slits (about 50\% in the outer parts). When using the photometric
redshifts, the targets were chosen within the redshift range $z\sim
0.6 - 1.3$. We used this broad interval due to the lack of $R$-band
data which causes a large uncertainty (degeneracy) in the location of
the 4000 \AA\ break for galaxies in the interval
$z\sim0.5-0.9$. Furthermore, the errors in redshift estimates are
$\sim0.15$ close to the magnitude limit of the sample.  In the case of
EIS0533-2412 only $IJK$ data were available.  In this case, the
targets were selected by searching for the expected $(I-K)$ and
$(J-K)$ colours of early-type galaxies in the redshift interval of
interest. In the case of EIS0046-2930, expected to be at lower
redshifts, galaxies were selected in the magnitude range of $19\lsim
I_{AB} \lsim22.5$, while for EIS0533-2412 and EIS0954-2023 they were
drawn in the interval $21\lsim I_{AB} \lsim23$.

A full description of the colour selection used to build the list of
spectroscopic targets likely to be cluster members will be presented
in a forthcoming paper (J{\o}rgensen \etal 2002).

\section{VLT Spectroscopy}
\label{sec:obs_red}

\begin{table*}[ht]
\center
\caption{Summary of the observations.}
\label{tab:obs}
\begin{tabular}{cccccccccc}
\hline\hline
Candidate & date & seeing & Inst. & Nr. of & Integration  & Observed & Measured & Stars & No  \\
        &          &  (arcsec)        & & masks & time & objects & redshifts & & Identification\\
\\
\hline
EIS0046-2930 & 24/09/2000 & 0.6-0.7 & FORS1 & 4   & 3600s & 85 & 63 & 5 & 17\\
EIS0533-2412 & 25/12/2000 & 0.7-1.2 & FORS2 & 1 & 14400s  & 47 & 30 & 4 & 13\\
EIS0954-2023 & 26/12/2000 & 0.5-0.7 & FORS2 & 2 & 14400s & 80 & 54 & 11 & 15 \\
\hline\hline
\end{tabular}
\end{table*}

The spectroscopic observations presented here were carried out using
FORS1 in the MOS mode (September 2000) on the VLT-ANTU telescope and
FORS2 in the MXU mode on the VLT-Kueyen telescope (December 2000)
(Cf. http://www.eso.org/instruments). In the MOS mode FORS1 provides
19 movable slit blade pairs that can be placed in the available
field-of-view. For the present observations Grism 150I+17 with the
order separation filter OG590 was used, providing a useful field of
$4.7 \times 6.8$ square arcmin and covering the spectral range
6000-11000\AA. The dispersion of 230\AA/mm (5.52\AA/pixel) yielded a
spectral resolution of 280 or about 29~\AA\ for a slit width of
1.4~arcsec.  The FORS2 observations were carried out in the
multi-object (MXU) spectroscopy mode using GRISM 200I+28 with order
separation filter OG550 covering the spectral range 5600-11000\AA. The
dispersion of 162~\AA/mm (3.89~\AA/pixel) yielded a resolution of 380
or about 21~\AA\ for a slit width of 1.2~arcsec. The slit mask allowed
for a much larger multiplex than was possible with FORS1 and typically
over 35~objects could be observed simultaneously.

The spectroscopic data were reduced using standard IRAF routines.  The
extracted one-dimensional spectra were then inspected to identify
spectral features and obtain a preliminary redshift estimate. The
spectra were then cross-correlated to template spectra using the FXCOR
task of IRAF. In general, the differences between the three redshift
estimates (emission lines, absorption features, and cross correlation)
were small and the redshift measured by the cross-correlation was
adopted.  A more detailed account of the reduction procedure will be
presented elsewhere (J{\o}rgensen \etal 2002).

A summary of the spectroscopic observations and the results obtained
from the data analysis is presented in Table~\ref{tab:obs}. Individual
exposure times ranged from 900 to 1800~sec depending on the total
integration time required for each mask. Note that in the case of
EIS0533-2412 two masks were prepared but technical problems prevented
the use of one of them at the time of the observation. In the case of
EIS0954-2023 one of the masks was 30~minutes shorter.  Altogether, a
total of 212 objects were observed yielding 147 measured redshifts in
the redshift interval 0.14-1.32.

\section{Results}
\label{results}

\begin{figure*}
\resizebox{!}{5.5cm}{\includegraphics{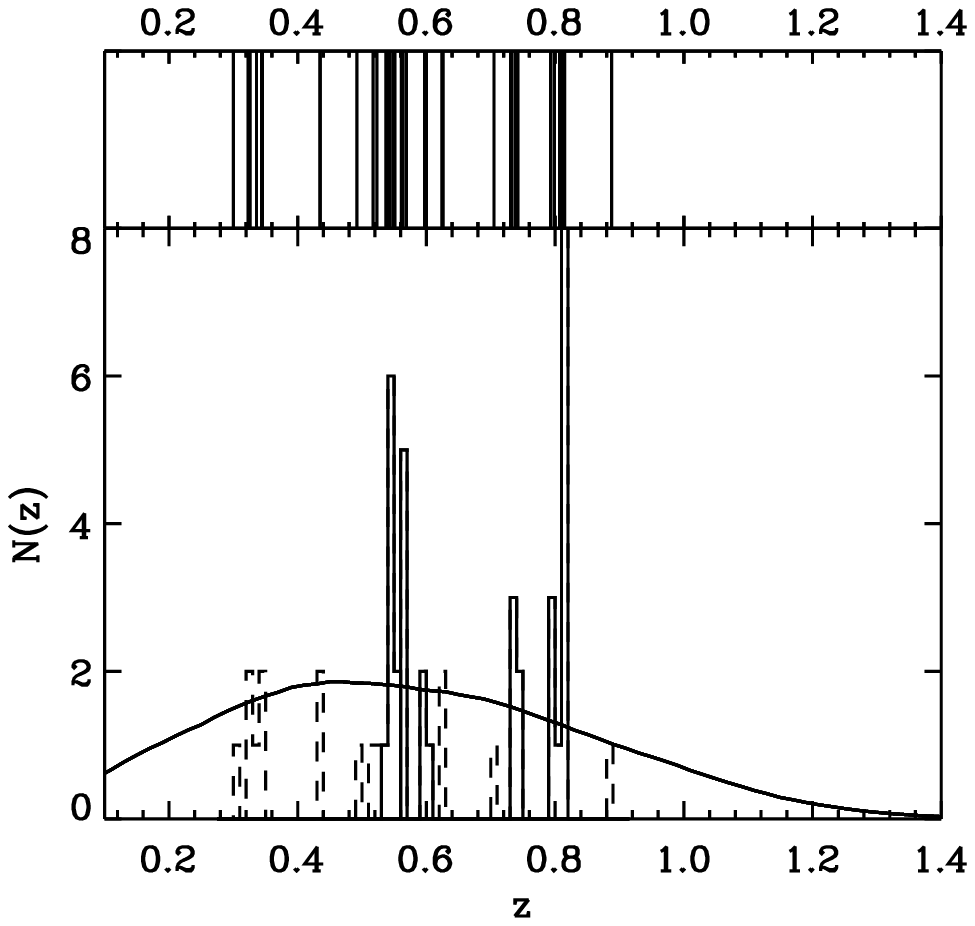}}
\hfill 
\resizebox{!}{5.5cm}{\includegraphics{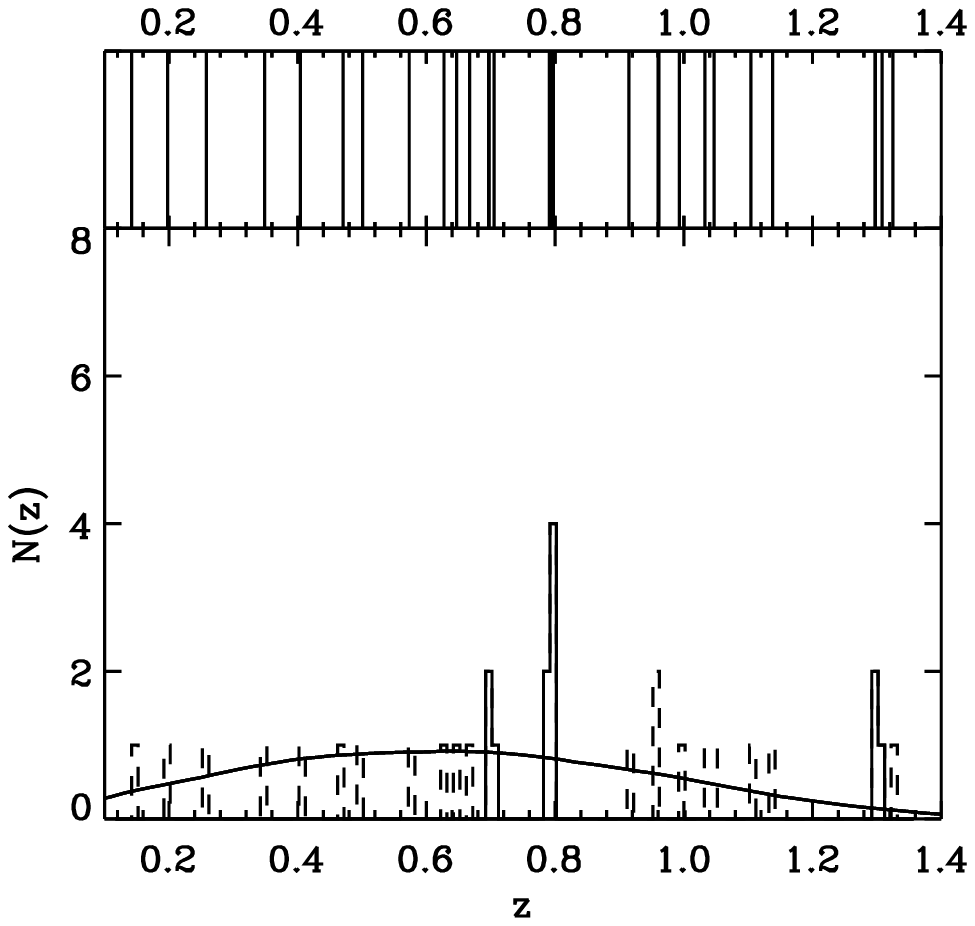}}
\hfill 
\resizebox{!}{5.5cm}{\includegraphics{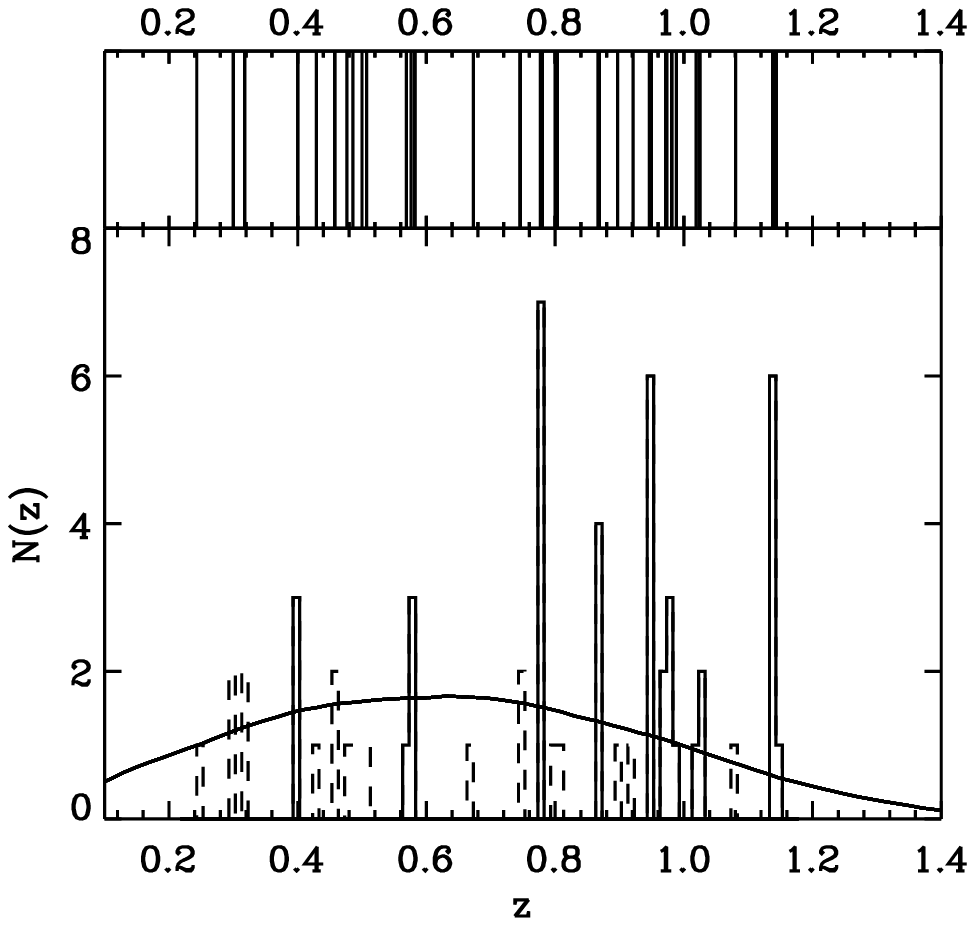}}
\caption{Redshift distribution of galaxies in the fields of
EIS0046-2930 (left), EIS0533-2412 (middle) and EIS0954-2023
(right). The upper part of the panels shows each individual
measurement while in the lower part the measurements are grouped in
bins $\Delta z= 0.01$ wide. The solid histograms indicate galaxy
groups identified in redshift space as described in the text.}
\label{fig:zhist_0046}
\end{figure*}

In Figure~\ref{fig:zhist_0046} we present the distribution of measured
redshifts for each of the fields considered showing in the upper part of
each panel the individual redshifts and in the lower part the redshift
distribution in bins $\Delta z= 0.01$ wide. The solid histograms
indicate groups that have been identified from the analysis of the
redshift distribution as discussed below. For each field the measured
redshift distribution is compared to that expected for a uniform
distribution of galaxies with a given luminosity function (LF) and
selected in the same magnitude intervals as the observed galaxies. The
$I$-band LF was computed using the LF parameters, split into three
spectral classes (early, spiral and late-types), as derived from the
$R$-band data of the ESO-Sculptor Survey (de Lapparent \etal 2002) up
to $z\sim0.6$. We assumed that the LF remains constant at higher
redshifts. The $I$-band LF at different redshifts was then obtained
using the appropriate SEDs for the three spectral classes considered
(Arnouts \etal 2002). It was confirmed that this approach leads to a
redshift distribution which is consistent with that measured by
the CFRS survey when the same limiting magnitude is adopted.  The
predicted distribution is normalised by requiring the number of
objects with $z>0.5$ to be equal to the number of galaxies
observed. This is done to approximately simulate the colour/photometric
redshift criteria adopted in selecting the target galaxies.

From the figure one can immediately see the presence of several peaks
in the observed distribution relative to the uniform background. Note
that the redshift range covered by the observations of the
EIS0046-2930 ($0.2<z<0.9$) is smaller than that for the other two
cases which have measured redshifts up to $z\sim1.3$.  This is due to
the brighter magnitude interval adopted in selecting the galaxy
sample. This also explains the difference in the predicted redshift
distribution for this field. It is also interesting to point out that
while most of the redshifts lie beyond $z\sim0.5$, as originally
intended, some have lower redshifts. The fraction of galaxies with
$z\lsim0.5$ is $\lsim$ 10\% of the total number of galaxies with
redshifts, nearly all corresponding to faint ($I_{AB}\sim 21-23$)
objects arbitrarily selected to fill-in available slits.
Typically about 50\% of the observed galaxy sample was 
arbitrarily selected, reflecting the fact that the area covered with
infrared data (limited by the size of the SOFI field) was too small to
position all the slits of the spectrograph.

\begin{figure*}
\center
\resizebox{!}{7.5cm}{\includegraphics{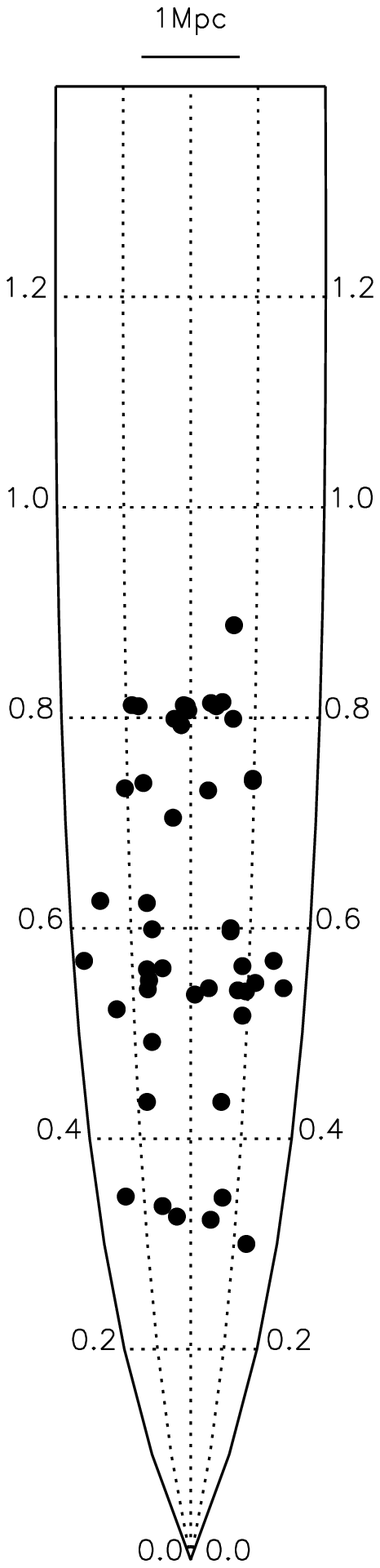}}
\resizebox{!}{7.5cm}{\includegraphics{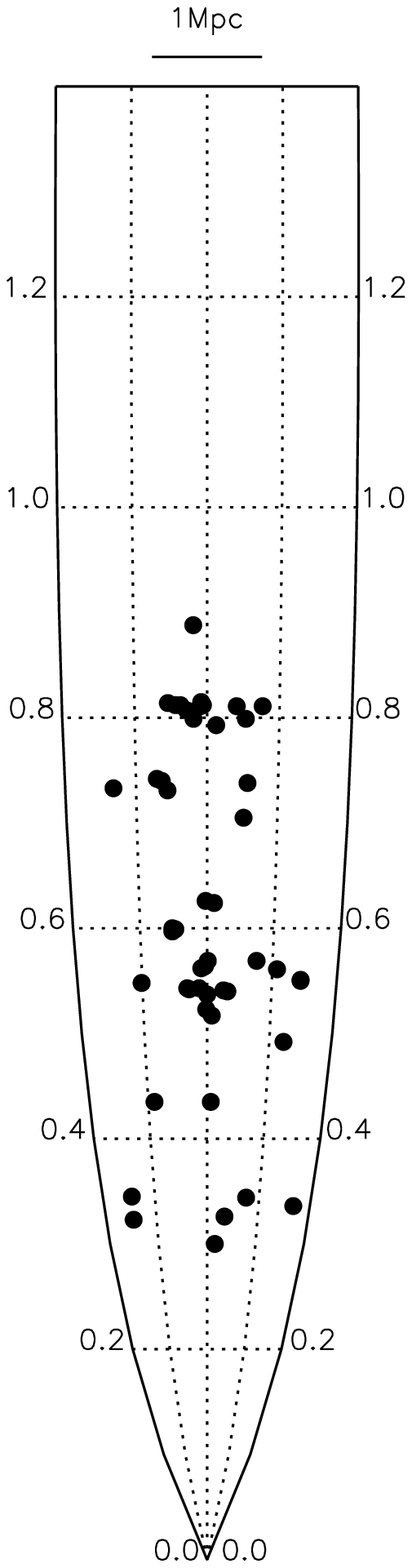}}
\hspace{1cm}
\resizebox{!}{7.5cm}{\includegraphics{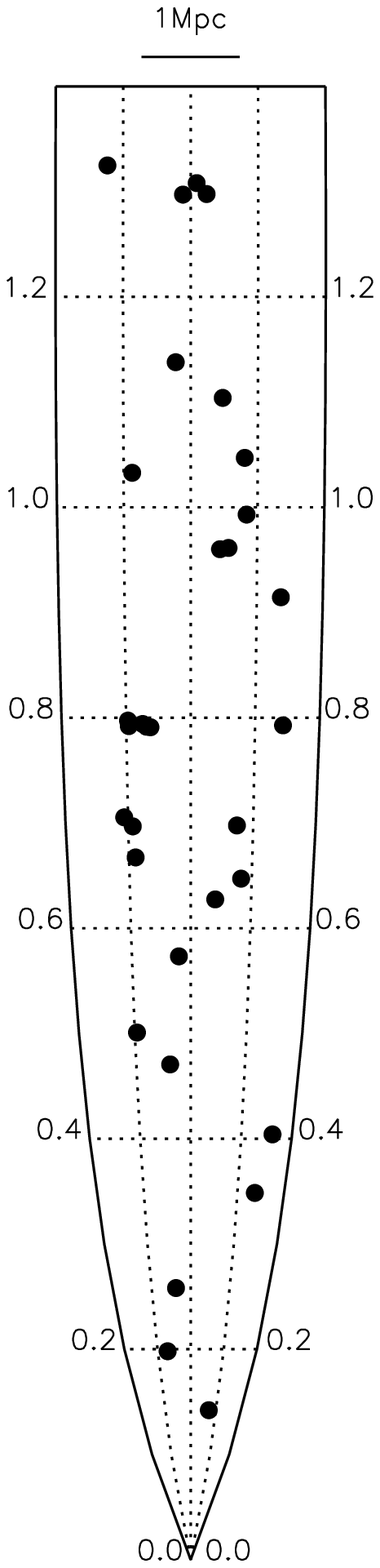}}
\resizebox{!}{7.5cm}{\includegraphics{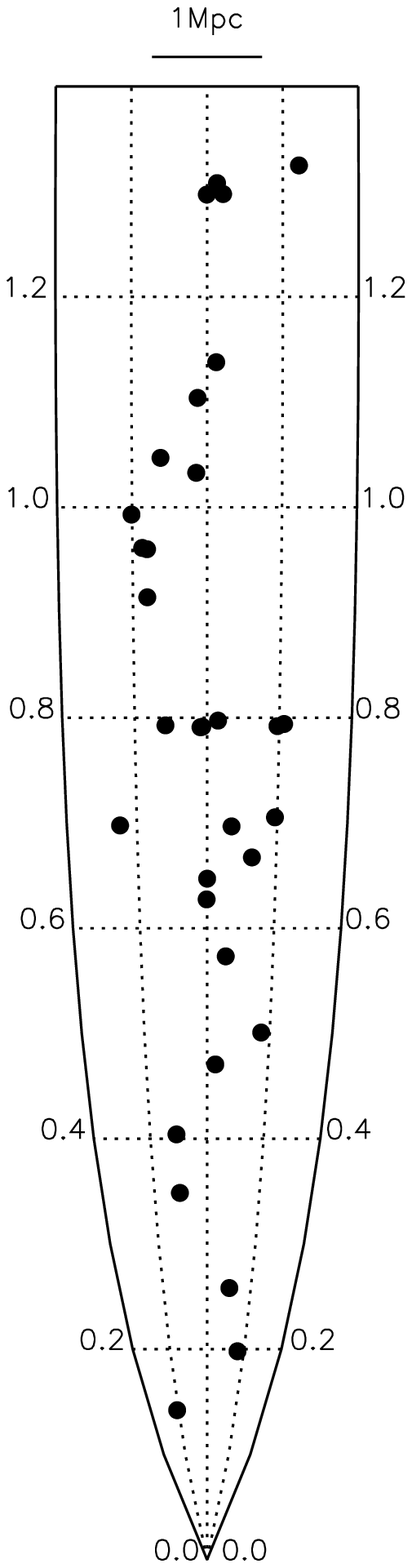}}
\hspace{1cm}
\resizebox{!}{7.5cm}{\includegraphics{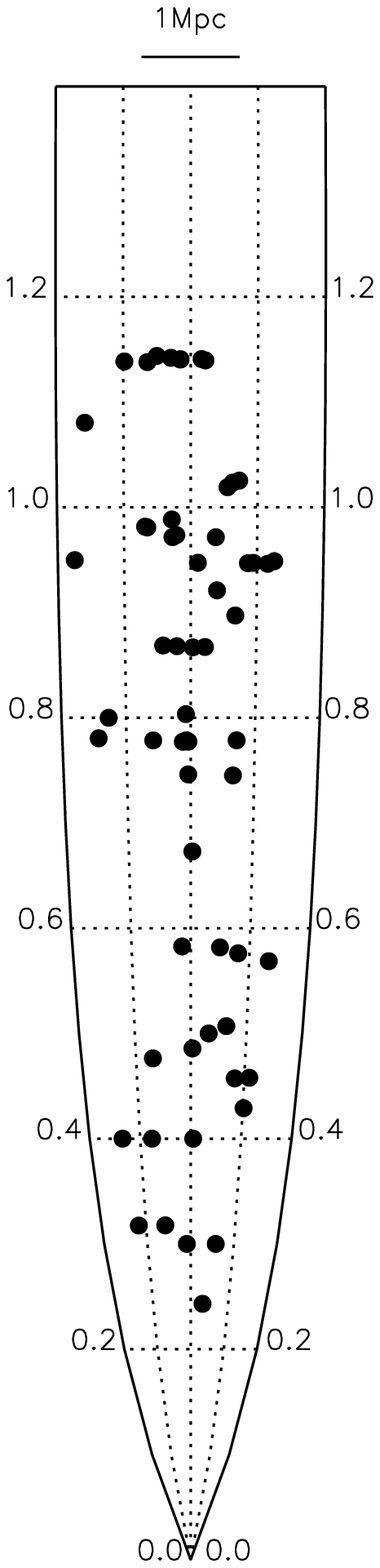}}
\resizebox{!}{7.5cm}{\includegraphics{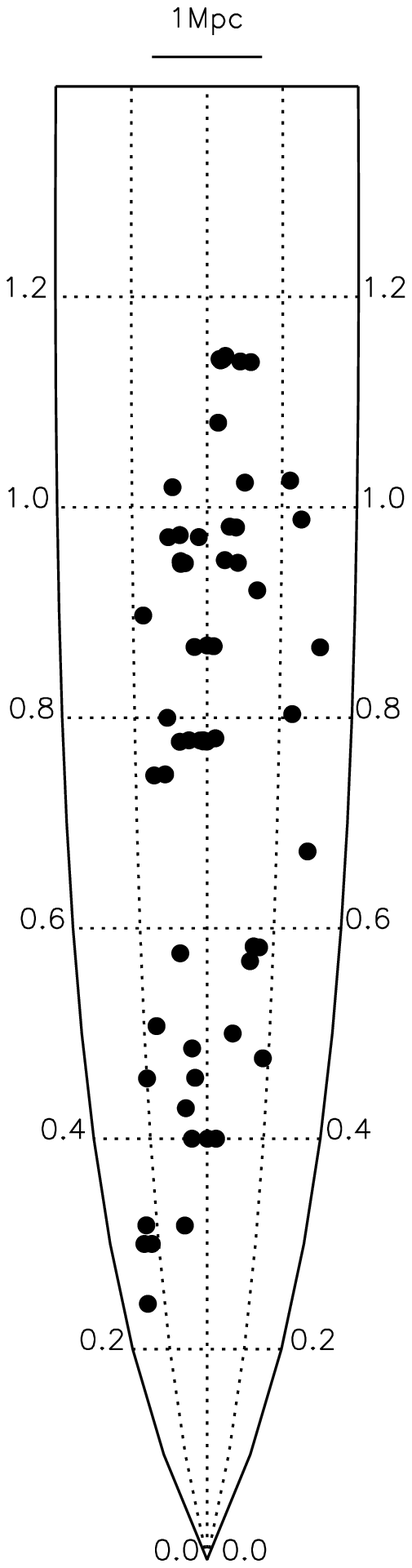}}
\caption{Redshift versus right ascension (left) and declination
(right) for the fields of EIS0046-2930, EIS0533-2412 and EIS0954-2023
corresponding to an angular scale of $10 \times 10$~arcmin. The scale of
1Mpc corresponds to $z=1.2$. The shape of the cones translates the
evolution of that scale with redshift.}
\label{fig:cones_00462930}
\end{figure*}

Given the incompleteness of the sample, groups have been identified
first in redshift space and then by their angular proximity. Groups in
redshift-space have been identified using the ``gap''-technique of
Katgert \etal (1996) which identifies gaps in the redshift
distribution larger than a certain size to separate individual
groups. In this preliminary analysis we have adopted a redshift gap of
$\Delta z=0.005*(1+z)$ corresponding to 1500~\kms in the rest-frame.
A total of five, three and eight groups with more than 3~members were
found in the fields of EIS0046-2930, EIS0533-2412 and EIS0954-2023,
respectively. To assess the significance of these detections we have
resorted to simulations and evaluated how frequently peaks similar to
those identified can occur by chance drawing galaxies with $z>0.5$
from a uniform background. We find that 8 groups (2 in EIS0046-2930, 2
in EIS0533-2412 and 4 in EIS0954-2023), corresponding to 50\% of the
groups identified, are likely to correspond to real enhancements in
redshift space ($99\%$ confidence level). In addition, we can also ask
how many of these are also spatially concentrated. This can be done by
examining Figure~\ref{fig:cones_00462930} which shows, for each field,
diagrams plotting redshifts as a function of right ascension and
declination. From the figure it is easy to identify the most
compelling cases of galaxies not only with concordant redshifts but
also within a circular region roughly 1~arcmin in radius. These cases
are listed in Table~\ref{tab:groups} which gives: in column (1) the
name of the cluster; in column (2) the ID of the group within each
field; in columns (3) and (4) the Right Ascension and the Declination;
in column (5) number of members; in columns (6) and (7) the mean
redshift and the standard deviation in \kms. We remind the reader that
all these cases are firm ($99\%$) detections in redshift space and
their location is in excellent agreement ($< $1~arcmin) with the
position of the original candidate as identified by the matched-filter
analysis.

\begin{table*}
\center
\caption{Groups identified in the cluster fields.}
\label{tab:groups}
\begin{tabular}{lrccrrc}
\hline\hline
Candidate & ID & Ra & Dec & $N_{g}$ & $\langle z\rangle$ & $\sigma$ (kms$^{-1}$)  \\
\hline
EIS0046-2930 &  1 & 00:46:29.6  & -29:30:57.4 & 12 & 0.808 & 1171 \\
\hline
EIS0533-2412  & 1 & 05:33:40.3  & -24:12:43.8 &  3 & 1.301 & -  \\
\hline
EIS0954-2023  & 1 & 09:54:47.5  & -20:23:55.2 &  6 & 0.948 & 202  \\
 & 2 & 09:54:37.0 & -20:22:54.7 & 8 & 1.141 & 285 \\
\hline\hline
\end{tabular}
\end{table*}

\begin{figure*}
\center
\resizebox{!}{8cm}{\includegraphics{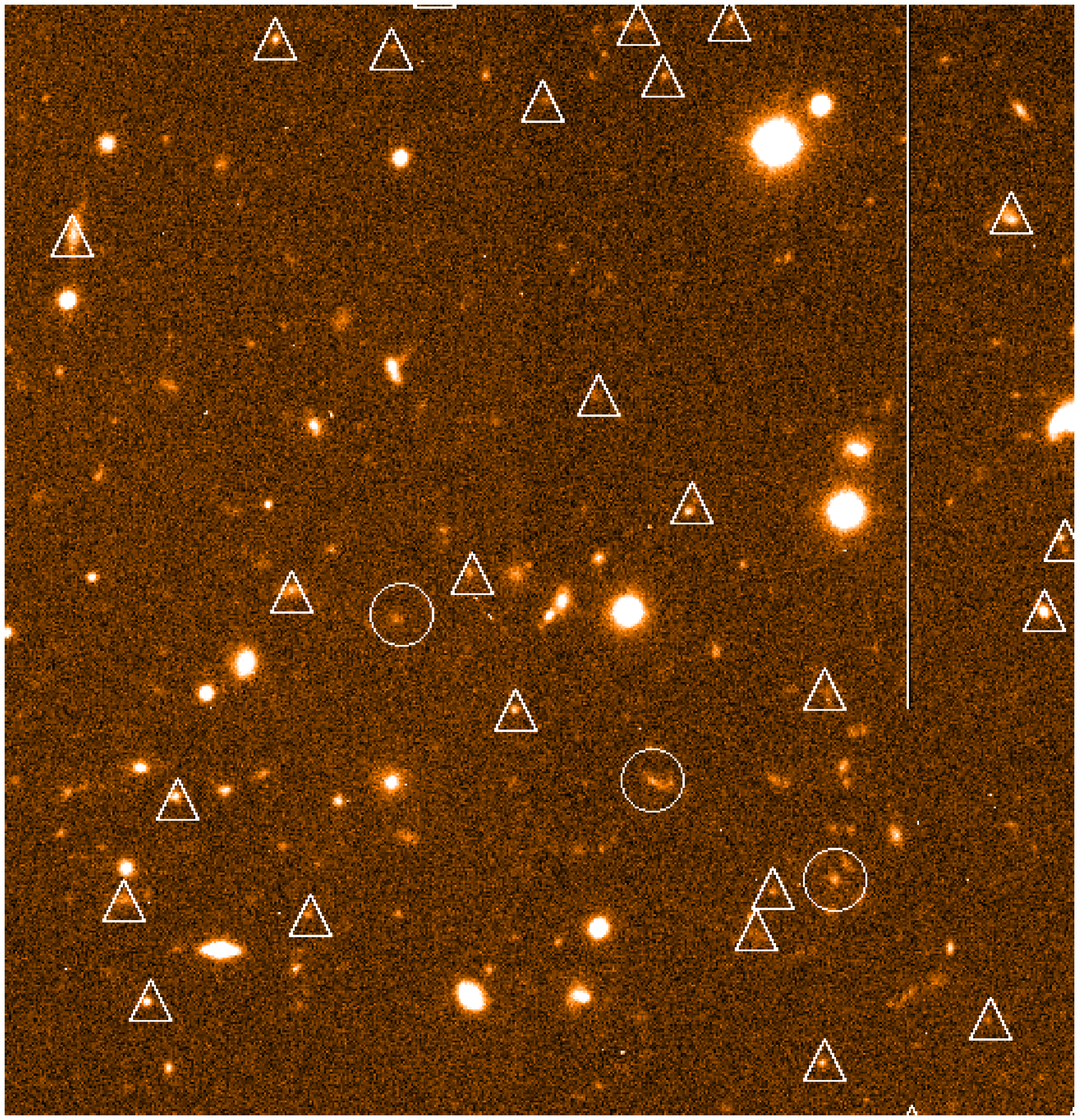}}
\caption{Image cutout 2~arcmin to the side taken from a FORS $R$ image
of the cluster field EIS0533-2412. Circles show galaxies with similar
spectroscopically confirmed redshift, while triangles those with
similar colours. Note the concentration of faint galaxies near the
lower right corner.}
\label{fig:images}
\end{figure*}

In summary, in all three candidate cluster fields considered we
identify at least one significant concentration of galaxies in
redshift and in position. In the field of EIS0046-2930 we identify a
system at $z=0.808$. This candidate was originally assigned a
matched-filter redshift of $z\sim0.6$ but subsequent work using
optical/infrared colour-magnitude diagrams (da Costa \etal 1999)
suggested it to be at higher redshift. Indeed, the apparently
brightest galaxy in the cluster was measured to be at $z=0.81$
(Ramella, private communication). New measurements of another 11
galaxies with concordant redshifts now corroborate this earlier
result. In the field of EIS0533-2412 we find a significant
concentration at $z=1.3$, which coincides with the location of the
matched-filter detection at an estimated redshift of $z=1.1$. While
currently only three galaxies have measured redshifts in the $z=1.3$
system, most of the faint galaxies in the field have similar colours as
shown in Figure~\ref{fig:images}. Furthermore, recent analysis of an
XMM-Newton image finds a $>5\sigma$ detection centred near the
location of the brightest cluster member for which we have a secure
redshift at $z=1.3$.  In fact, the distribution of galaxies with
similar $(J-K)$ colour extends some 3~arcmin to the NE, where another
X-ray detection has been found. A more detailed discussion of the
X-ray data will be presented elsewhere (Neumann \etal 2002). Finally,
in the field of EIS0954-2023 we find evidence for two clumps. One at
$z=1.141$, at the same location as the matched-filter detection, and
the other a foreground concentration at $z=0.95$, some 2~arcmin away
from the original detection.  High resolution cutouts for the systems
discussed here can be found at the URL
``$http::www.obs-nice.fr/benoist/high-z.clusters.html$''.

\section{Summary}
\label{summary}

This paper presents new spectroscopic data of EIS cluster candidate
fields identified from moderately deep $I$-band images using the
matched-filter algorithm. The three fields considered were selected
because the cluster candidates had estimated redshifts beyond $z=0.8$.
Analysis of the spectroscopic data strongly suggests the existence of
real density enhancements at high redshifts ($0.8<z<1.3$) in all of
them. The measured redshifts are, in general, consistent with those
estimated from the photometric data. In at least one of the cluster
fields, the location of the high-redshift system coincides remarkably
well with a robust X-ray detection, lending further support to the
reality of the system.  Therefore, despite the fact that it is
difficult to decide if those clumps are filaments or bound systems,
for two of them the evidence seem to favour the second possibility:
for the $z=0.81$ clump a red sequence is observed (da Costa 
\etal 1999) whereas the $z=1.3$ one has an X-ray detection associated 
with it.

The present paper strongly suggests that the EIS cluster candidate
catalog provides a valuable pool from which to construct a statistical
sample of optically-selected clusters at high redshift. The present
data alone contribute with four systems at $z>0.8$ in the southern
hemisphere, two of which at $z\gsim1$, ideal for VLT studies. The
success in identifying significant concentrations from a relative
small sample underscores the importance of collecting multi-band
optical/infrared data and estimating photometric redshifts to select
potential cluster members. However, in establishing the true nature of
these systems will require a better sampling of these systems which
will become possible with the availability of an integral field unit
as foreseen by the VIMOS spectrograph.

\acknowledgements

We would like to thank the EIS Team for the effort of producing the
publicly available object catalogs for the EIS-Wide and Pilot
Surveys. LFO thanks the SARC and Carlsberg Foundations for financial
support during the project period.


{}

\end{document}